# Analyzing Walter Skeat's Forty-Five Parallel Extracts of William Langland's *Piers Plowman*

## Roger Bilisoly[1]

[1]Department of Mathematical Sciences, Central Connecticut State University, 1615 Stanley St, New Britain, CT 06050-4010

**Abstract**

Walter Skeat published his critical edition of William Langland's 14th century alliterative poem, *Piers Plowman*, in 1886. In preparation for this he located forty-five manuscripts, and to compare dialects, he published excerpts from each of these. This paper does three statistical analyses using these excerpts, each of which mimics a task he did in writing his critical edition. First, he combined multiple versions of a poetic line to create a best line, which is compared to the mean string that is computed by a generalization of the arithmetic mean that uses edit distance. Second, he claims that a certain subset of manuscripts varies little. This is quantified by computing a string variance, which is closely related to the above generalization of the mean. Third, he claims that the manuscripts fall into three groups, which is a clustering problem that is addressed by using edit distance. The overall goal is to develop methodology that would be of use to a literary critic.

**Key Words:** Text mining, Edit distance, Clustering, Randomization tests, Middle English, Alliterative poetry

## 1. Introduction

When writing a literary text, the author's original drafts often differ from the various editions of his or her work. This may be due to an editor making changes, or by unintentional modifications introduced in the process of publishing the text. One ambitious undertaking by a literary critic is to construct a critical edition of a text, where the goal is to produce a version closest to the author's intentions given the existing versions along with whatever auxiliary knowledge exists.

The original drafts of more recent authors often survive, but as one goes back in history, this becomes less common. For medieval texts, the surviving manuscripts copied by scribes are typically all that exist today. For example, the alliterative poem, *Sir Gawain and the Green Knight*, is contained in only one manuscript, which is a scribal copy, and nothing is known about the identity of its author.

This paper uses statistical ideas to try to reproduce three aspects of the process that Walter Skeat, a 19th century historical linguist and medievalist, used when he wrote a critical edition of William Langland's *Piers Plowman*. These techniques are built on a generalization of the mean and variance developed by Fréchet and Pennec for Riemannian manifolds that turns out to work for other distance functions. In the applications below, edit distance on strings is used.

## 2. Defining a Mean and Variance Based on Edit Distance

Xavier Pennec developed a theory of doing statistical testing of data from a Riemannian manifold, which is described in Pennec (1999) and Pennec (2006). He notes that Maurice Fréchet generalized the mean and variance in the 1940s (published in French), but he did not develop it further. The key insight is defining the mean as a minimizing value of a sum of squared distances, which is outlined below.

### 2.1 Fréchet's Generalization of the Mean and Variance

The usual mean can be defined as the value, $c$, that minimizes $f(c)$ as defined in Equation (2.1), and this minimum value is the variance.

$$f(c) = \frac{1}{n} \sum_{i=1}^{n} |x_i - c|^2 \qquad (2.1)$$

Because $|x_i - c|$ is the Euclidean distance between $x_i$ and $c$, replacing it with an arbitrary distance function, $d$, is a straightforward generalization, which is shown in Equation (2.2). Moreover, sums of squares are widespread in statistics, so this is an appealing approach.

$$f(c) = \frac{1}{n} \sum_{i=1}^{n} d(x_i, c)^2 \qquad (2.2)$$

For Fréchet and Pennec, $d$ is the geodesic distance for a Riemannian manifold. A simple application of this is given in Section 2.2 of Bilisoly (2013), where the data are angles (that is, *circular data*, in the sense of Fisher (1996)). This paper uses Equation (2.2) with the Levenshtein edit distance, but other string metrics could be used, such as the ones given in Elzinga (2006), which outlines a general approach to constructing such metrics for categorical time series. One complication with using a non-Euclidean metric, $d$, is that uniqueness need not hold.

Equation (2.2) can be further generalized. For instance, although sum of squares are ubiquitous in statistics, other powers could be used, as noted in Bilisoly (2013). The most important example is using the first power: minimizing Equation (2.1) with respect to $c$ then gives the median, and the resulting minimum of $f(c)$ is the mean absolute deviation. The author has experimented with this, but it is not used here because minimizing Equation (2.3) is more prone to non-uniqueness than (2.2).

$$f(c) = \frac{1}{n} \sum_{i=1}^{n} d(x_i, c) \qquad (2.3)$$

### 2.2 Levenshtein's Edit Distance

Levenshtein (1966) introduced the idea of a distance between pairs of strings. The idea is to define it as the minimum number of edits needed to transform one into the other. Levenshtein used three types: insert one character, delete one character, and substitute one character for another. Although other edits have been considered in the literature, these three are still popular. It turns out that this distance makes strings a metric space. Section 3 of Bilisoly (2013) gives an example of the computing the algorithm (which is an example of dynamic programming) for the distance between "old" and "halde" (two Middle English spellings of "old" from McIntosh et al. (1986)). The distance is at most three because one can substitute $h$ for $o$, insert an $a$ and an $e$, and the algorithm guarantees that there is no shorter way to do this.

The above algorithm is widely implemented. For example, Mathematica has the function `EditDistance[]`, which is also able to handle the additional letters that appear in Middle English. For this poem, some manuscripts use the characters yogh and thorn, which are written ȝ and þ, respectively. For example, these letters appear in line 67 of manuscript I of Skeat (1885) in the phrase "ȝe þat beoþ mene." Note that the numbering used here and below is from this reference.

When minimizing Equation (2.2), one has to decide over which set of values of $c$ the optimization considers. In this context there are two possibilities. First, one could optimize over all strings, which allows the possibility that the mean (understood to be the Fréchet mean below) could be a word or phrase that has never been found in the existing literature. In historical linguistics, reconstructed words are of interest. For example, Proto-Indo-European, the hypothetical ancestor to all the Indo-European languages, is entirely reconstructed because it predates writing (see the first eight chapters of Fortson (2010).) Second, in work with manuscripts, there is much interest in *attested* spellings, which means they actually appear in at least one text (as is done in McIntosh et al. (1986)). The latter approach is more natural for this application, so by combining edit distance with Pennec's work, and by optimizing only over the words found in Skeat (1885), the analyses can begin.

## 3. The Fréchet Mean and Skeat's Critical Edition

In preparation for his critical edition of William Langland's poem, *Piers Plowman*, Skeat tried to compile a complete list of manuscripts containing this work, which eventually grew to a total of forty-five. According to the introductory notice of Skeat (1885), he "carefully examined every manuscript personally, with only two exceptions," from which he concluded that there were three distinct versions of the poem, the A-, B- and C-texts, not two as was earlier believed. Finally, he published Skeat (1886a), which contains his best reconstructions, one for each of the three versions, along with Skeat (1886b), which has background information, notes, and indexes.

Among other things, Skeat was an expert in Middle English dialects. He thought that Langland was born at Cleobury Mortimer, Shropshire, England, which is still accepted today. Consequently, one of his goals was to reject spellings from other dialects introduced by scribes that lived and worked elsewhere. Pages 7-24 of Skeat (1885) explicitly remarks on the quality of the manuscripts, so he certainly does not think they are all equally good. For example, he says of manuscript XII, it "is much corrupted in places, whilst *the dialect has been turned into Northumbrian*" (the italics are his.) He, however, thinks manuscript XIII is the best overall: "I still adhere to my opinion that it may indeed be the author's autograph copy." So if Skeat were willing to give each manuscript a weight based on its overall quality, these would be far from uniform.

For a numeric variable, there are many ways to compute a typical value of a sample. For example, the mean or the median are commonly used, and if the quality of the observations vary, one might use a weighted mean, perhaps trimmed. The goal of this section is to construct the best poetic lines by computing the mean using edit distance for each word of a given line. As Skeat did, we will only work with one of the A-, B-, or C-texts at a time. Unlike him, however, we assume that the manuscripts are equally useful because even crudely estimating weights would require Middle English dialect expertise that this author does not have.

## 3.1 Reconstructing One Line Using the Fréchet Mean

Skeat wrote three versions of *Piers Plowman*, the A-, B-, and C-texts. In this section, using the forty-five Passus III extracts from Skeat (1885), using the mean, we reconstruct line 70 of from the A-texts, line 79 from the B-texts, and line 80 from the C-texts, as numbered in Skeat (1886a). Note that these three line numbers refer to the same poetic line, the one that mentions brewers, bakers, butchers, and cooks.

> Brewesteres and bakesteres bocheres and cokes
> Brewsters and baksteres bochers and cokus
> Brewsteres and bakesteres bocheres and kokes
> Brewesters and Baxsters bochers and Cookes
> brewsters & baksters bochers and kokes
> Brewesters and baxters bowchers and cookes
> Brewsteris and baksteris bocheris and cokis
> Brewsters and baksteris bocheris and cokis
> brewsteris & bakesteris bocheris  and cokes
> Breusters & bakesters bochers & cokis
> brousters & bakers bocheris and cokis
> Brusterrs and bakesters bochers and cokes
> Brusters and bakesters bouchers and cokes
> Brewesters and  Baksters Bochiers and Cokes
> Brewesters and baksters bochers & cokes
> Boþe websteres & bakesterys & bocheres & Cookys
> Brewers and bakers bochers and kokes

**Figure 1:** The seventeen versions of line 79, Passus III, of the B-texts.

Although this is one of the less variable poetic lines, Figure 1 displays much spelling variation among the seventeen B-texts: in fact, no two of these lines are identical. In addition, this line alliterates, meaning that it uses repeated initial sounds. Here the letter *b* is used in three of the four stressed words, the so-called aa/ax stress pattern, which is quite common in Middle English alliterative poems, which is quantified in Oakden (1968).

Conceptually, creating the optimal version of this line is straightforward. Ignoring capitalization, one finds the mean of the seventeen forms of "brewers," again for "bakers," "butchers,"  "cooks," and if one likes, for "and." The result is (3.1).

> Brewsters and baksters bochers and cokes       (3.1)

Skeat was an expert in Middle English dialects, so optimizing at the level of words makes sense. However, he did not think all the manuscripts were equally reliable. In fact, he preferred the first B-text over all others, and that, (3.2), was the line he used in his critical edition.

> Brewesteres and bakesteres bocheres and cokes     (3.2)

Instead of words, one could optimize over entire lines. For the sake of comparison, the mean of the seventeen lines in Figure 1 is given by (3.3), which turns out to be the third to last line in Figure 1. This, however, is much closer to (3.1) than Skeat's (3.2).

$$\text{Brewesters and baksters bochers \& cokes} \tag{3.3}$$

For the A-texts, a complication arises. The mean of the fourteen forms of "cooks" is not unique, a possibility noted in Section 2.1. The two means are "cokes," which is the mode, and "kokes." The two reconstructions are (3.4) and (3.5), and this differs from Skeat's (3.6), which is the same as the line in his first A-text, the one that he believes is the most accurate manuscript of that version.

$$\text{Brewers and bakers bochers and cokes} \tag{3.4}$$

$$\text{Brewers and bakers bochers and kokes} \tag{3.5}$$

$$\text{Brewesters bakers bochers and cookes} \tag{3.6}$$

Finally, for the C-texts another counterintuitive result happens. For "brewers" the spelling "brewers" appears ten times, and the following forms each appear once: "breweres," "breueres," "bruweres," and "brueres." One would guess that the mean would be "brewers," but it is, in fact, "breweres." This happens because the distances are squared in Equation (2.2), and edit distance between "brewers" and "brueres" is 3, while "breweres" has distance of 1 with all but one of the other forms, the exception having distance of 2. However, if Equation (2.3) were used, then "brewers" would be answer, which makes sense because it is more like the median. For comparison to the above, the C-text reconstruction is (3.7), while Skeat's is (3.8), which is the same as the first C-text, the one that he thinks is best of that version.

$$\text{Bakers and breweres bochers and cokes} \tag{3.7}$$

$$\text{As bakers and brewers bouchers and cokes} \tag{3.8}$$

Clearly equal weighting of texts is not working here because Skeat judges some manuscripts to be much more accurate than others. However, many lines only appear in lower quality versions, and he wants his critical edition to be as complete as possible. Moreover, even his favorite manuscripts have errors, so he did combine texts, but not in a fashion that is similar to the above mean computation.

## 4. Edit Distance Variance and Skeat's Variability Claim

In the introduction of Skeat (1886b), he discusses many issues such as the dating of the three versions of *Piers Plowman* and the original dialect of the poem. For the quantitatively minded, the claim on page x is notable: "I may here remark that the manuscripts of the B-text agree, in general, very closely, and that the number of various readings is small." That is, of the three versions, the B-texts are the least variable, and the goal of this section is to quantify this.

The forty-five manuscript excerpts published in Skeat (1885) all refer to the same narrative event, but there is much textual variation. First, the A- and B-texts generally have eleven lines, while the C-texts only have nine lines. However, there are five exceptions: IX is missing the last line; XX is missing the seventh and eighth lines; XXVIII is missing the end of sixth line and the beginning of the seventh; XLII has two extra lines; and XLIII has one extra line. These five excerpts were excluded from the analysis below because the edit distance is sensitive to differences in string length.

Another complication is that not all the lines in the C version correspond well to the A and B versions. To get the best matches, the analysis below uses only the third, fourth, and the last four lines of the forty remaining excerpts. The Appendix gives Skeat's reconstructions from his critical edition to allow the reader judge the merit of this decision.

Once the decision on which lines and which excerpts is made, then the process of finding the mean by minimizing Equation (2.2) also gives the variance, which is the value of this minimum. Table 1 gives the variances of the three versions, and as Skeat predicts, the B-texts are the least variable.

| | |
|---|---|
| A-texts | 4338.38 |
| B-texts | 2189.93 |
| C-texts | 2713.33 |

**Table 1:** The Fréchet variances for Skeat's three manuscript types of *Piers Plowman*. Excerpts with missing or extra lines are excluded.

That Table 1 agrees with Skeat's claim could be due to chance, and to see how likely the variance of the B-texts is the smallest, a randomization test is performed under the null hypothesis that all excerpts come from the same population. Specifically, the analysis that led to Table 1 is repeated 5000 times, where each repetition randomly permutes the order of the excerpts. Two test statistics are used here, Var(A)/Var(B) and Var(C)/Var(B), which, being a ratio of variances, mimics the F statistic, as noted in Bilisoly (2014). The value of these two ratios for the original order is 1.981 and 1.239, and Figure 2 shows the results. There are three points that satisfy Var(A)/Var(B) > 1.981 and Var(C)/Var(B) > 1.239, so the empirical p-value is 3/5000, and we reject the null hypothesis. That is, Skeat's claim that the B-texts are the least variable is probably true.

During his life, Skeat tried to locate every text of *Piers Plowman*, and in doing so, he increased the number of known manuscripts from twenty-nine to forty-five. Certainly he wanted to study the population of the surviving copies of this poem, and he probably found most of them. For example, SEENET (2014) says there are over fifty unique versions known today, which is not much larger than Skeat's forty-five. This brings up a statistical issue: should one worry about over-fitting models or not? This may be a rare case where the peculiarities of the sample are of interest to literary critics, so over-fitting is acceptable.

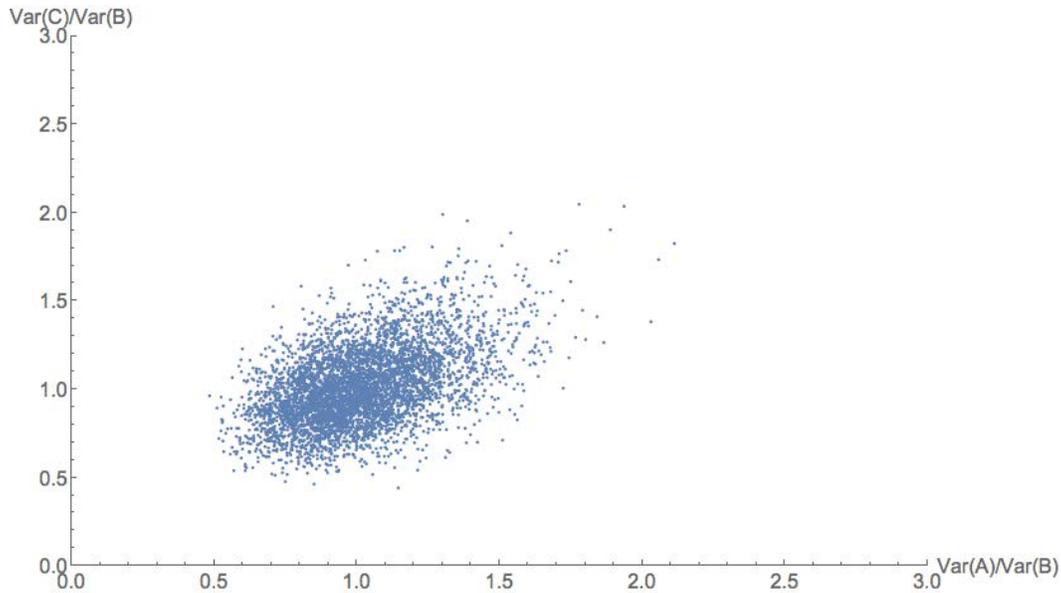

**Figure 2:** A plot of variance ratios for 5000 randomizations of the order of the forty excerpts. Only three points satisfy Var(A)/Var(B) > 1.981 and Var(C)/Var(B) > 1.23.

## 5. Edit Distance for Clustering

Skeat's critical edition convinced scholars that there were three versions of *Piers Plowman*, not two, which is an example of clustering the manuscripts, using today's terminology. Using edit distance, the analyses in Sections 3 and 4 already compute the distance matrix, which is all one needs to do clustering, and this section compares the result of the algorithm to Skeat.

Mathematica's `FindCluster[]` was used with `DistanceFunction` set to `EditDistance` and `Method` set to `"Optimize"`. Using the six lines described above for all forty-five excerpts, the results most comparable to Skeat's were obtained for four clusters: the first with 11 As; the second 3 As, 7 Bs, and 1 C; the third 10 Bs, and 2 Cs; the last 11 Cs. Because Skeat has both domain expertise and he analyzed entire manuscripts (ranging from roughly 2,000 to 7,000 lines) versus using only six lines here, the agreement here seems remarkably good.

## 6. Conclusions

The results of Section 3 show that Skeat did not treat the manuscripts as equally valuable in making his critical edition, which he clearly states in Skeat (1885). Section 4 quantifies his claim that the B-texts are least variable, while Section 5 supports Skeat's claim of there being more than two versions of *Piers Plowman*. So it seems that quantitative tools could supplement human expertise when working with multiple versions of a text. Skeat himself might have agreed considering that his first academic job was a lectureship in mathematics at Cambridge.

# Appendix

The excerpts below are from Skeat (1886a), which is his critical edition of *Piers Plowman*. The top is the A-text, the middle the B-text, and the bottom the C-text.

Meires and maistres and ȝe that beoth mene
Bitwene the kyng and the comuns to kepe the lawes
As to punisschen on pillories or on pynnyng stoles
Brewesters bakers bochers and cookes
For theose be men vppon molde that most harm worchen
To the pore people that percel-mel buggen
Thei punisschen the peple priueliche and ofte
And recheth thorw regratorie and rentes hem buggeth
With that the pore people schulde puten in heore wombe
For toke thei on trewely thei timbrede not so hye
Ne bouȝte none borgages beo ȝe certeyne

Meires and maceres that menes ben bitwene
The kynge and the comune to kepe the lawes
To punyschen on pillories and pynynge stoles
Brewesteres and bakesteres bocheres and cokes
For thise aren men on this molde that moste harme worcheth
To the pore peple that parcel-mele buggen
For they poysoun the peple priueliche and oft
Thei rychen thorw regraterye and rentes hem buggen
With that the pore people shulde put in here wombe
For toke thei on trewly thei tymbred nouȝt so heiȝe
Ne bouȝte non burgages be ȝe ful certeyne

ȝut Mede myldeliche the meyre hue bysouhte
Bothe shereues and seriauns and suche as kepeth lawes
To punyshen on pillories and on pynyng-stoles
As bakers and brewers bouchers and cokes
For thees men doth most harme to the mene puple
Richen thorw regratrye and rentes hem byggen
With that the poure puple sholde putten in hure womben
For toke they on triweliche they tymbrid nat so heye
Nother bouhten hem burgages be ȝe ful certayn

# References

Bilisoly, R. (2013). "Generalizing the Mean and Variance to Categorical Data Using Metrics," *American Statistical Association Proceedings of the Joint Statistical Meetings*, Section on Statistical Learning and Data Mining, American Statistical Association, Alexandria, Virginia, 2266-2275.

Bilisoly, R. (2014). "Quantifying Prosodic Variability in Middle English Alliterative Poetry," *American Statistical Association Proceedings of the Joint Statistical Meetings*, Section on Statistical Learning and Data Mining, American Statistical Association, Alexandria, Virginia, 1230-1241.

Elzinga, C. (2006). "Sequence Analysis: Metric Representations of Categorical Time Series." Downloaded from ResearchGate on September 22, 2015, https://www.researchgate.net/profile/Cees_Elzinga/publication/228982046_Sequence _analysis_Metric_representations_of_categorical_time_series/links/5464a15e0cf2c0c 6aec64294.pdf.

Fisher, N. I. (1993). *Statistical Analysis of Circular Data*, Cambridge University Press, Cambridge, UK.

Fortson IV, B. W. (2010). *Indo-European Language and Culture, An Introduction*, 2$^{nd}$ edition, Wiley-Blackwell, Malden, Massachusetts.

Levenshtein, V. I. (1966). "Binary Codes Capable of Correcting Deletions, Insertions, and Reversals," *Soviet Physics Doklady*, 10 (8), 707-710.

McIntosh, A., Samuels, M. L., and Benskin, M. (1986). *A Linguistic Atlas of Late Mediaeval English*, 4 Volumes, Aberdeen University Press, Aberdeen, UK.

Oakden, J. P. (1968). *Alliterative Poetry in Middle English: The Dialectal and Metrical Survey*, Archon Books.

Pennec, X. (1999). "Probabilities and Statistics on Riemannian Manifolds: Basic Tools for Geometric Measurements," *International Workshop on Nonlinear Signal and Image Processing*, Antalya, Turkey.

Pennec, X. (2006). "Intrinsic Statistics on Riemannian Manifolds: Basic Tools for Geometric Measurements," *Journal of Mathematical Imaging and Vision*, 25(1), 127-154.

Skeat, W. W. (1885). *Parallel Extracts from Forty-Five Manuscripts of Piers Plowman*, Second Edition, Truebner.

Skeat, W. W. (1886a). The Vision of William Concerning Piers the Plowman in Three Parallel Texts, Volume I, Oxford.

Skeat, W. W. (1886b). The Vision of William Concerning Piers the Plowman in Three Parallel Texts, Volume II, Oxford.

SEENET, The Society for Early English and Norse Electronic Texts (2014). http://piers.iath.virginia.edu/index.html. Accessed 09/28/2015.